\documentclass[11pt,a4paper]{article}
\usepackage{graphicx}
\usepackage{here}
\usepackage{caption}
\usepackage{tabularx}
\usepackage{amsmath,amssymb}
\usepackage{bm}
\usepackage{wrapfig}
\usepackage{color}
\usepackage{braket}
\usepackage{enumerate}
\newcommand{\be}{\begin{equation}}
\newcommand{\ee}{\end{equation}}

\addtocounter{section}{0}
\makeatother
\title{Correlation functions of harmonic lattices in d-dimensional space}
\author{
Masafumi Shimojo \\
{\it Department of Electronics and Information Engineering, }\\
{\it National Institute of Technology, Fukui college} \\
{\it Geshicho, Sabae-Shi, Fukui 916-8507, Japan} \\
\\
Satoshi Ishihara 
 \footnote{E-mail:satoshi@hyogo-u.ac.jp},  \ \ 
Hironobu Kataoka,
Atsuko Matsukawa,\ \ Kazuo Koyama \\
{\it Department of Physics, Hyogo University of Education} \\
{\it Shimokume, Kato-shi, Hyogo 673-1494, Japan} 
}
\begin{document}
\maketitle
\begin{abstract}
We study the correlation functions between the dynamical variables and between their conjugate momenta at sites of a harmonic lattice in the $d$-dimensional Euclidean space. We show that at the thermodynamic limit, they can be expressed in terms of Lauricella's C-type hypergeometric functions. Furthermore, using these expressions, we explicitly demonstrate that the correlators near the center of the lattice satisfying Dirichlet boundary conditions coincide with those for the lattice with the periodic boundary conditions. By utilizing these expressions, we expect to make it easier to create programs that compute fast and highly precise for the quantum information quantities of subsystems within lattices.
\end{abstract}
\section{Introduction}
Indices of information loss due to entanglements, such as R\'{e}nyi entropy are subjects of research not only in quantum information theory but also in many other fields, including field theory and condensed matter physics\cite{Amico}. In the harmonic lattice model representing the discretized Klein Gordon field, these indices are determined from the vacuum correlators $Q_{\vec{r}\vec{s}}=\braket{q_{\vec{r}}q_{\vec{s}}}$ and $P_{\vec{r}\vec{s}}=\braket{p_{\vec{r}}p_{\vec{s}}}$ for the set of sites of interest, where $q_{\vec{r}},q_{{\vec{s}}}$ and $p_{\vec{r}},p_{\vec{s}}$ are quantized dynamical variables of the sites with position vectors $\vec{r},\vec{s}$  and their conjugate momentum operators, respectively. Hereafter, we will simply refer to these correlators as $Q$ and $P$ or $Q$,$P$ matrices as needed. 

For example, consider an isolated lattice $U$ and its subset $A$ of sites, where we evaluate 
R\'{e}nyi entropy $S_A^{(n)}$, between $A$ and its complement.  The procedure is as follows: First, compute the matrices  $Q_{\vec{r}\vec{s}}$ and $P_{\vec{r}\vec{s}}$ restricted to sites $\vec{r}$ and $\vec{s}$ belonging to $A$. Second, find the eigenvalues $\xi_i^2$ of $\Xi = QP$. The R\'{e}nyi entropy of degree $n$ can be computed using the following formula\cite{Nobili}.
\be
S_A^{(n)}   
= \frac{1}{n-1}\sum_{i=1}^{\dim(\mathcal{H}_A)}
\log\left( (\xi_i+\frac{1}{2})^n - (\xi_i-\frac{1}{2})^n\right), \label{Renyi2}
\ee
where $\mathcal{H}_A$ is the Hilbert space of the subsystem $A$ and $n\geq 2$ is an integer parameter. 

For the computations on finite lattices to approximate those in the thermodynamic limit, the lattice spacing 1, the size $l_A$ of the subsystem $A$, and the entire lattice size $L$ must satisfy the constraint $1 \ll l_A \ll L$.
On the other hand, as $l_A$, $L$, and the spatial dimension $d$ increase, computation of the matrices $Q$ and $P$ becomes extremely time-consuming. 

In this manuscript, we show that by considering the thermodynamic limit, the lattice scale $L\to\infty$ and replacing the sum over all site positions $(k_1, k_2,\ldots, k_d)$ within $U$ appearing in the computation of $Q$ and $P$ with multiple integrals $\int\int\dots\int dk_1 dk_2 \dots dk_d$ and calculating them analytically, the results for any spatial dimension are expressed by the C-type Lauricella's hypergeometric functions, $F_C$, irrespective of the boundary conditions for the dynamical variables $q_{\vec{r}}$s and their conjugate momenta $p_{\vec{r}}$s. We expect that by utilizing these expressions along with appropriate numerical computation libraries, it makes easier to create programs that rapidly compute various information quantities arising from entanglement.
\section{Correlators on finite harmonic lattices}
We consider a free scalar field in $(d+1)$ dimensional Lorentzian space-time and discretize it into a cubic lattice with $L^d$ sites of harmonic oscillators coupled with nearest neighborhood. The Hamiltonian is given by 
\begin{equation}
H = \frac{1}{2}\sum_{\vec{j}}\left(p_{\vec{j}}^2+\omega^2q_{\vec{j}}^2 +\sum_{\alpha =1}^d(q_{\vec{j}+\vec{e}_\alpha}-q_{\vec{j}})^2\right), \label{Hamil2}
\end{equation}
where $\vec{j}=(j_1,j_2,\ldots,j_{d})$ denotes the position of a site and $\vec{e}_\alpha=(0,0,\ldots,0,1,0,\ldots,0)$ is a unit vector in the $\alpha$ direction.

When the lattice $U$ satisfies periodic boundary conditions(PBCs) along directions of spatial axes,  
$q_{\vec{j}+L\vec{e}_{\alpha}}=q_{\vec{j}}, \ p_{\vec{j}+L \vec{e}_{\alpha}}=p_{\vec{j}}
,\ \alpha=1,2,\ldots, d$, solving Hamilton's canonical equations yields the variables $q_{\vec{j}}$s and $p_{\vec{j}}$s. Quantizing these then, the dispersion relation $\omega_k$ and $Q$, $P$ matrices are given \cite{IK2MS_1} 
by
\begin{align}
\omega_{\vec{k}}^{(P)} & = \sqrt{\omega^2 + 4\sum_{\alpha=1}^d \sin^2\frac{\pi k_\alpha}{L}}\ , \label{omegakp} \\
Q_{\vec{r}\vec{s}}=\braket{q_{\vec{r}}q_{\vec{s}}} &  = \frac{1}{2V}\sum_{\vec{k}} \frac{1}{\omega_{\vec{k}}^{(P)}}\prod_{\alpha=1}^d \cos \frac{2\pi k_\alpha(r_\alpha -s_\alpha )}{L},
\label{corpq}
\\
P_{\vec{r}\vec{s}}=\braket{p_{\vec{r}}p_{\vec{s}}} & = \frac{1}{2V}\sum_{\vec{k}}\omega_{\vec{k}}^{(P)}\prod_{\alpha=1}^d \cos \frac{2\pi k_\alpha(r_\alpha-s_\alpha)}{L}, \label{corpp}
\end{align}
where $k_\alpha $s are integers such that $0\leq k_\alpha \leq L-1$ and $V=L^d$. 

In the case where $U$ satisfies Dirichlet boundary conditions(DBCs) as follows:
\begin{align*}
q_{(j_1,j_2,\ldots, j_\alpha=L,\ldots,j_{d})}  & =q_{(j_1,j_2,\ldots,\ j_{\alpha}=0,\ \ldots,j_{d})}=0 , \\
\ p_{(j_1,j_2,\ldots, j_\alpha=L,\ldots,j_{d})} & =p_{(j_1,j_2,\ldots,\ j_\alpha=0\ ,\ldots,j_{d})}=0,
\end{align*}
the dispersion relation and $Q$, $P$ matrices are given \cite{IK2MS_1} by:
\begin{align}
\omega_{\vec{k}}^{(D)}  & = \sqrt{\omega^2 + 4\sum_{\alpha=1}^d \sin^2\frac{\pi k_\alpha}{2L}}, \label{omegakf} \\
Q_{\vec{r}\vec{s}}=\braket{q_{\vec{r}}q_{\vec{s}}} 
& = \frac{1}{2V}\sum_{\vec{k}}\frac{1}{\omega_{\vec{k}}^{(D)}}
\prod_{\alpha=1}^d 2\sin\left(\frac{\pi r_\alpha k_\alpha}{L}\right)\sin\left(\frac{\pi s_\alpha k_\alpha}{L}\right), \label{corfq}\\
P_{\vec{r}\vec{s}}=\braket{p_{\vec{r}}p_{\vec{s}}} 
& = \frac{1}{2V}\sum_{\vec{k}}\omega_{\vec{k}}^{(D)}
\prod_{\alpha=1}^d 2\sin\left(\frac{\pi r_\alpha k_\alpha}{L}\right)\sin\left(\frac{\pi s_\alpha k_\alpha}{L}\right), \label{corfp}
\end{align}
where, as in the case of PBCs, $V=L^d$. Note that, as long as the size $L$ of $U$ is finite,
for any spatial axis direction, $j_\alpha$ ranges 0 to $L$ and the number of $j_\alpha $s is $L+1$, but truly dynamical $q_{(j_1,j_2,\ldots,j_\alpha,\ldots,j_{d})}$ and $p_{(j_1,j_2,\ldots,j_\alpha,\ldots,j_{d})}$ exist only at $1\leq j_\alpha \leq L-1$ and $k_\alpha$s also take integers in the range $1\leq k_\alpha\leq L-1$. 
\section{Correlation functions in the thermodynamic limit} 
In this section, we show that the correlation functions $Q_{\vec{r}\vec{s}}$ in (\ref{corpq}),(\ref{corfq}) and $P_{\vec{r}\vec{s}}$ in (\ref{corpp}), (\ref{corfp}) are both expressed by the same type of hypergeometric series in the thermodynamic limit, regardless of the boundary conditions.  
In order to discuss common properties beyond differences in $Q$ and $P$, we also denote $Q_{\vec{r}\vec{s}}$ as $G^{(1/2)}_{\vec{r}\vec{s}}$, $P_{\vec{r}\vec{s}}$ as $G^{(-1/2)}_{\vec{r}\vec{s}}$ for the case that the entire lattice satisfies PBCs and $J_{\vec{r}\vec{s}}^{(\pm 1/2)}$ for the lattice satisfying DBCs. 

First, in correlators in the finite lattice satisfying PBCs, (\ref{corpq}) and (\ref{corpp}), taking the limit as $L\to \infty$ , and replacing $2\pi k_\alpha/L \to k_\alpha$, $\sum_{\vec{k}}\to\int_0^{2\pi}\dots\int_0^{2\pi} L^d/(2\pi)^dd^dk $
yields the following expression:
\be
G^{(\pm 1/2)}_{\vec{r}\vec{s}}  =\frac{1}{2(2\pi)^d}\int_0^{2\pi}\dots \int_0^{2\pi}d^dk\ (\omega_k^{(P)})^{\mp 1}
\prod_{\alpha=1}^d \cos n_\alpha k_\alpha,  \label{Gp}
\ee
where $n_\alpha$ denotes $|r_\alpha - s_\alpha|$. 
The dispersion relation and its reciprocal are expressed by
\begin{align}
& (\omega_k^{(P)})^{\pm 1}  =(\omega^2+2\sum_{\alpha=1}^d (1-\cos k_\alpha))^{\pm 1/2} =(\Omega)^{\pm 1} (1-v\sum_{\alpha=1}^d \cos k_\alpha)^{\pm 1/2} \nonumber \\
& = \Omega^{\pm 1}\sum_{m=0}^\infty
  \begin{pmatrix}
       \pm 1/2\\
        m
  \end{pmatrix} (-v)^m \left(\sum_{\alpha}^d \cos k_\alpha\right)^m 
=\Omega^{\pm 1}\sum_{m=0}^\infty
  \frac{(\mp 1/2)_m}{m!} v^m \left(\sum_{\alpha=1}^d \cos k_\alpha\right)^m,  \label{omega2}
\end{align}
where we set $\Omega^2=\omega^2+2d$, $v=2/\Omega^2$, and on the right-hand side of the last equality, we use the Pochhammer symbol$(a)_m=a(a+1)\dots (a+m-1)$. Substituting (\ref{omega2}) into (\ref{Gp}), we obtain 
\be
G^{(\pm 1/2)}_{\vec{r}\vec{s}}  =\frac{\Omega^{\mp 1}}{2(2\pi)^d}\sum_{m=0}^\infty
  \frac{(\pm 1/2)_m}{m!} v^m 
\int_0^{2\pi}\dots \int_0^{2\pi}d^dk\ 
\left(\sum_{\alpha=1}^d \cos k_\alpha\right)^m\prod_{\alpha'=1}^d \cos n_{\alpha'} k_{\alpha'}. 
\label{GP2}
\ee
Here, we denote the multiple integral appearing in (\ref{GP2}) as $I_{m,n_1,n_2,\ldots,n_d}$. 
Then it can be expressed as a linear combination of the products of integrals with respect to each $k_\alpha$  by expanding $(\sum_{\alpha=1}^d \cos k_\alpha)^m$ as follows:
\be
I_{m,n_1,\dots,n_d} 
=\sum_{l_1=0}^{m}\sum_{l_2=0}^{m}\dots\sum_{l_d=0}^{m} \delta_{m,(\sum_{\alpha'=1}^d l_{\alpha'})}
\frac{m!}{l_1!l_2!\cdots l_{d-1}!l_d! }\left(\prod_{\alpha=1}^d I_{l_\alpha,n_\alpha}\right), \label{Imnd}
\ee
where $I_{l_\alpha,n_\alpha}$ is given by
\be
I_{l_\alpha,n_\alpha}=\int_0^{2\pi}\cos^{l_\alpha}k_\alpha \cos n_\alpha k_\alpha dk_\alpha=
 \begin{cases}
 \displaystyle \frac{2\pi}{2^{l_\alpha}} 
 \begin{pmatrix}
  l_\alpha \\
   (l_\alpha-n_\alpha)/2
 \end{pmatrix} & \text{if}\ l_\alpha \geq n_{\alpha}\ \text{and}\ l_\alpha=n_\alpha\hspace{-2mm} \mod 2, \\
  0 & \text{otherwise}.
 \end{cases} \label{Ialpha}
\ee
Putting $(l_\alpha-n_{\alpha})/2\overset{\underset{\mathrm{def}}{}}{=}p_\alpha$, we have:
\be
\frac{m!}{l_1!l_2!\cdots l_{d-1}!l_d! }
   \prod_{\alpha=1}^d 
    \begin{pmatrix}
     l_\alpha \\
     (l_\alpha-n_\alpha)/2
    \end{pmatrix}
= \frac{m!}{\prod_{\alpha} p_\alpha! (n_\alpha+p_\alpha)!}
= \frac{m!}{\prod_{\alpha} p_\alpha! n_\alpha!(n_\alpha+1)_{p_\alpha}}. \label{mlll}
\ee
Furthermore, when we put $N\overset{\underset{\mathrm{def}}{}}{=} n_1+n_2+\dots+n_d$, $p_1+p_2+\dots +p_d=(m-N)/2\overset{\underset{\mathrm{def}}{}}{=} u $ holds. 
Let $a=\pm 1/2$, then we obtain the following equation:
\be
\left(\pm \frac{1}{2}\right)_m=(a)_{N+2u}=(a)_N(a+N)_{2u}=(a)_N2^{2u}\left(\frac{a+N}{2}\right)_u\left(\frac{a+1+N}{2}\right)_u. \label{a1_2}
\ee
Using (\ref{Imnd}),(\ref{Ialpha}),(\ref{mlll}) and (\ref{a1_2}), the equation (\ref{GP2}) can be written as 
\be
G^{(\pm 1/2)}_{\vec{r}\vec{s}}
=\frac{\Omega^{\mp 1}}{2}\frac{(a)_N}{\prod_{\alpha} n_\alpha!}\frac{v^N}{2^N}
   \sum_{p_1=0}^\infty \sum_{p_2=0}^\infty \dots\sum_{p_{d}=0}^\infty
   \frac{\left(\frac{a+N}{2}\right)_u\left(\frac{a+1+N}{2}\right)_u}{\prod_\alpha(n_\alpha+1)_{p_\alpha}}
   \frac{v^{2u}}{\prod_\alpha p_\alpha !}\ . \label{GP3}
\ee
The infinite series appearing in equation (\ref{GP3}) is Lauricella's C-type hypergeometric series defined by:
\be
F^{(d)}_C(\alpha,\beta ;\gamma_1,\gamma_2,\dots,\gamma_d;z_1,z_2,\dots,z_d)=\sum_{p_1=0}^\infty \sum_{p_2=0}^\infty \dots\sum_{p_{d}=0}^\infty 
\frac{(\alpha)_u(\beta)_u}{\prod_{\alpha=1}^d (\gamma_\alpha)_{p_\alpha}}
\frac{\prod_{\alpha=1}^d z_\alpha^{p_\alpha}}{\prod_{\alpha=1}^d p_\alpha !}\ ,
\ee
and the $Q$ and $P$ correlation functions are expressed, respectively, as follows:
\begin{align}
Q_{\vec{r}\vec{s}}^{(P)} & = \frac{1}{2\Omega}\frac{\left(\frac{1}{2}\right)_N}{\prod_\alpha n_\alpha !}\frac{v^N}{2^N} 
 F_C^{(d)}\left(\frac{N+1/2}{2},\frac{N+3/2}{2};n_1+1,n_2+1,\ldots,n_d+1;v^2,v^2,\ldots,v^2\right), \label{QP_L} \\
P_{\vec{r}\vec{s}}^{(P)} & = \frac{\Omega}{2}\frac{\left(-\frac{1}{2}\right)_N}{\prod_\alpha n_\alpha !}\frac{v^N}{2^N}
F_C^{(d)}\left(\frac{N-1/2}{2},\frac{N+1/2}{2};n_1+1,n_2+1,\ldots,n_d+1;v^2,v^2,\ldots,v^2\right). \label{PPL}
\end{align}
The convergence condition for $F_C^{(d)}(\alpha,\beta;\gamma_1,\ldots,\gamma_d;v^2,v^2,\dots, v^2$) appearing in the expressions (\ref{QP_L}) and (\ref{PPL}) is $v\cdot d<1$ or $v\cdot d=1 \wedge\ \gamma_1+\gamma_2+\dots + \gamma_d-\alpha-\beta>0$. Since $v=2/(\omega^2+2d)$, in the massive case, the former condition is always satisfied. Even in the massless case, $\omega^2=0$, since $(n_1+1)+\dots+(n_d+1)-(a+N)/2-(a+1+N)/2=d-a-1/2$, the latter convergence condition is satisfied for $d\geq 2$.

Next, we consider the thermodynamic limit of the correlators, (\ref{corfq}) and (\ref{corfp}) in the lattice under DBCs. Replacing $\pi k_\alpha/L \to k_\alpha$, $\sum_{\vec{k}}\to \int_0^\pi \dots\int_0^\pi L^d/\pi^d d^dk $
 yields the following expression:
\be
J_{\vec{r}\vec{s}}^{(\pm 1/2)} =\frac{1}{2\pi^d}\int_0^{\pi}\dots \int_0^{\pi}d^dk (\omega_k^{(D)})^{\mp 1}
\prod_{\alpha=1}^d (\cos n_0^{(\alpha)} k_\alpha - \cos n_1^{(\alpha)} k_\alpha),  \label{Jrs0}\\
\ee
where 
\be
n_{\epsilon_\alpha}^{(\alpha)}
=\begin{cases}
 |r_\alpha - s_\alpha| ,\ \text{if\ } \epsilon_\alpha=0, \\
 r_\alpha + s_\alpha,\ \text{if\ } \epsilon_\alpha=1.  
\end{cases}\label{nepsilon}
\ee
Hereafter, unless confusion arises, the superscript $(\alpha)$ will be omitted. The dispersion relation is given by the following equation, as in (\ref{omega2}):
\be
\omega_{k}^{(D)}  = \sqrt{\omega^2+4\sum_{\alpha=1}^d\sin^2 k_\alpha/2}=\Omega \sqrt{1-v\sum_{\alpha=1}^d \cos k_\alpha}, 
  \text{\ \ \ where\ \ \ }\Omega^2= \omega^2+2d,\ v=\frac{2}{\Omega^2}. \label{omegaD}
\ee
Expanding $\prod_{\alpha=1}^d (\cos n_0^{(\alpha)} k_\alpha - \cos n_1^{(\alpha)} k_\alpha)$ yields the following expression:
\begin{align}
J_{\vec{r}\vec{s}}^{(\pm 1/2)} & = \sum_{\epsilon_1=0}^1\sum_{\epsilon_2=0}^1\dots \sum_{\epsilon_d=0}^1
   (-1)^{\sum_{\alpha=1}^d \epsilon_\alpha}J^{(\pm 1/2)}_{n_{\epsilon_1},n_{\epsilon_2},\dots,n_{\epsilon_d}},  \label{Jrs}\\
   & J^{(\pm 1/2)}_{n_{\epsilon_1},n_{\epsilon_2},\dots,n_{\epsilon_d}} 
   =\frac{1}{2\pi^d}\int_0^{\pi}\dots \int_0^{\pi}d^dk (\omega_k^{(D)})^{\mp 1}
      \prod_{\alpha=1}^d \cos n_{\epsilon_\alpha}  k_\alpha. \label{Jint}
\end{align}
The differences between (\ref{Gp}) and (\ref{Jint}), besides $n_\alpha$ changing to $n_{\epsilon_\alpha}$, are that (i) the coefficient before the integral changes from $1/(2\cdot 2^d\pi^d)$ to $1/(2\pi^d)$ and that (ii) the integration range for each $k_\alpha$ changes from $[0,2\pi]$ to $[0,\pi]$. 
Among these, due to (ii), the value of the integral for each $k_\alpha$ in (\ref{Jint}) becomes half that of (\ref{Ialpha}), and the entire multiple integral becomes multiplied by $ 1/2^d$. This cancels exactly with the change due to (i), so (\ref{Jint}) matches (\ref{Gp}) except for the substitution of $n_{\epsilon_\alpha}$ for $n_\alpha$. 
Thus, Eq.(\ref{Jint}) leads to
\be
J^{(\pm 1/2)}_{n_{\epsilon_1},n_{\epsilon_2},\dots,n_{\epsilon_d}}
=\frac{\Omega^{\mp 1}}{2}\frac{(a)_N}{\prod_{\alpha} n_{\epsilon_\alpha}!}\frac{v^N}{2^N}
   \sum_{p_1=0}^\infty \sum_{p_2=0}^\infty \dots\sum_{p_{d}=0}^\infty
   \frac{\left(\frac{a+N}{2}\right)_u\left(\frac{a+1+N}{2}\right)_u}{\prod_\alpha(n_{\epsilon_\alpha}+1)_{p_\alpha}}
   \frac{v^{2u}}{\prod_\alpha p_\alpha !}, \label{Jint3}
\ee
where we set $N\overset{\underset{\mathrm{def}}{}}{=} n_{\epsilon_1}+n_{\epsilon_2}+\dots+n_{\epsilon_d}$, $u\overset{\underset{\mathrm{def}}{}}{=} p_1+p_2+\dots p_d$ as before.
Then correlation functions $Q$ and $P$ under DBCs can be expressed, respectively, using Lauricella's hypergeometric functions as follows:
{\small
\begin{align}
& \hspace{-0.3cm} Q_{\vec{r}\vec{s}}^{(D)}  \nonumber \\
& \hspace{-0.3cm} =\sum_{\epsilon_1=0}^1\dots \sum_{\epsilon_d=0}^1(-1)^{\sum_\alpha\epsilon_\alpha}\frac{1}{2\Omega}\frac{\left(\frac{1}{2}\right)_N}{\prod_\alpha n_{\epsilon_\alpha} !}\frac{v^N}{2^N}
 F_C^{(d)}\left(\frac{N+1/2}{2},\frac{N+3/2}{2};n_{\epsilon_1}+1,\ldots,n_{\epsilon_d}+1;v^2,\dots,v^2\right),
  \label{QC_L}\\
& \hspace{-0.3cm} P_{\vec{r}\vec{s}}^{(D)}  \nonumber \\
& \hspace{-0.3cm}  = \sum_{\epsilon_1=0}^1\dots \sum_{\epsilon_d=0}^1(-1)^{\sum_\alpha\epsilon_\alpha}
\frac{\Omega}{2}\frac{\left(-\frac{1}{2}\right)_N}{\prod_\alpha n_{\epsilon_\alpha} !}\frac{v^N}{2^N}
F_C^{(d)}\left(\frac{N-1/2}{2},\frac{N+1/2}{2};n_{\epsilon_1}+1,\dots,n_{\epsilon_d}+1;v^2,\dots,v^2\right). \label{PCL}
\end{align}
\normalsize
Thus, we have shown that for harmonic lattices in space with $d\geq 2$, the correlation functions 
in the thermodynamic limit are expressed by hypergeometric functions of the same type, regardless of the boundary conditions of the lattice. 

Reference \cite{Coser} derives the necessary $Q$ and $P$ matrices in the thermodynamic limit to discuss the contour functions for the entanglement entropies in one-dimensional harmonic lattice.  By setting $d=1$ in the $Q$ and $P$ matrices given in (\ref{QP_L}), (\ref{PPL}) and (\ref{QC_L}), (\ref{PCL}), and performing some calculations, we obtain all the correlation functions in the appendix C of \cite{Coser}. 
\section{\large Connection between correlators of different boundary conditions}
In the thermodynamic limit, near the center of the lattice, $r_\alpha \sim s_\alpha \sim L/2$, the physical states are expected to be not largely affected by boundary conditions, and the correlation functions under the DBCs (\ref{QC_L}) and (\ref{PCL}) should coincide with those under PBCs (\ref{QP_L}) and (\ref{PPL}).
On the other hand, for $J^{(\pm 1/2)}_{n_{\epsilon_1},n_{\epsilon_2},\dots,n_{\epsilon_d}}$ of (\ref{Jint3}), when all $n_{\epsilon_\alpha}=n_0=|r_\alpha-s_\alpha| $, it coincides precisely with the PBC correlators (\ref{GP3}). Therefore, we examine the behavior of the other terms appearing in (\ref{Jrs}), (\ref{QC_L}) and (\ref{PCL}) in the thermodynamic limit, i.e,  the behavior of terms $J^{(\pm 1/2)}_{n_{\epsilon_1},n_{\epsilon_2},\dots,n_{\epsilon_d}}$ with some $n_{\epsilon_\alpha} =n_1=r_\alpha+s_\alpha\sim L\to \infty$. 
Since $J^{(\pm 1/2)}_{n_{\epsilon_1},n_{\epsilon_2},\dots,n_{\epsilon_d}}$ is symmetric under permutations of $n_{\epsilon_\alpha}$, setting $\epsilon_1=\epsilon_2=\dots=\epsilon_{d_1}=1$, $\epsilon_{d_1+1}=\dots = \epsilon_d=0$, we do not lose generality. So, we set $n_1^{(1)} + n_1^{(2)} + ... + n_{1}^{(d_1)} \overset{\underset{\mathrm{def}}{}}{=}  N_+$ and $N_- \overset{\underset{\mathrm{def}}{}}{=}  N - N_+$. Hereafter in this section, $n_{\epsilon_\alpha}^{(\alpha)}$ in formulae will be denoted by $n_\alpha$ for simplicity.

We evaluate the behavior of each factor in (\ref{Jint3}) as $n_1^{(\alpha)}\to \infty$. Using Stirling's formula, we obtain 
\begin{align}
(a)_N & \sim (N_+ +N_-+a)^{N_+ + N_- +a-1/2}e^{-(N_++N_-+a)}\sim N_+^{N_+}\left(1+\frac{N_-+a}{N_+}\right)^{N_+}N_+^{N_-+a-1/2}e^{-N_+ -N_-} \nonumber \\
& \sim N_+^{N_+}e^{-N_+}N_+^{N_-+a-1/2}. \label{aN}  
\end{align} 
For $\alpha=1,2,\dots, d_1$, let $n_{\epsilon_\alpha}^{(\alpha)} = \lambda_\alpha N_+$, we obtain the following expression: 
\be
 (\prod_{\alpha=1}^d n_\alpha !)^{-1}  \sim (e^{-N_+}\prod_{\alpha=1}^{d_1} n_\alpha^{n_\alpha+1/2})^{-1} 
\sim (\prod_{\alpha=1}^{d_1} \lambda_\alpha^{\lambda_\alpha N_+})^{-1} N_+^{-N_+} N_+^{-d_1/2} e^{N_+}. \label{ni!}
\ee
The formula for the maximum Shannon entropy of a information source with $d_1$ variables whose probability distribution is $(\lambda_1, \lambda_2,\dots, \lambda_{d_1})$ yields the following inequality: 
\be
\log (\prod_{\alpha=1}^{d_1} \lambda_\alpha^{\lambda_\alpha N_+})^{-1}
=N_+\sum_{\alpha=1}^{d_1}(-\lambda_\alpha \log \lambda_\alpha)\leq N_+\log d_1=\log d_1^{N_+}. \nonumber
\ee
This applies to (\ref{ni!}) and we obtain 
\be
 (\prod_{\alpha=1}^d n_\alpha !)^{-1} \leq d_1^{N_+} N_+^{-N_+} N_+^{-d_1/2} e^{N_+}. \label{ni!2}
\ee
Using (\ref{aN}) and (\ref{ni!2}) and $v =2/(\omega^2 + 2d)$, the evaluation of the coefficient of the infinite series in (\ref{Jint3}) is as follows:
\be
\frac{\Omega^{\mp 1}}{2}\frac{(a)_N}{\prod_\alpha n_{\alpha} !}\frac{v^N}{2^N}
\sim \left(\frac{d_1/d}{2}\right)^{N_+} O(N_+^{N_-+a-(1+d_1)/2}).  \label{Cofbehave}
\ee

Evaluation of one term in the hypergeometric series yields the following expression:
\be
\frac{\left(\frac{a+N}{2}\right)_u\left(\frac{a+1+N}{2}\right)_u}{\prod_\alpha(n_\alpha+1)_{p_\alpha}}
\frac{v^{2u}}{\prod_\alpha p_\alpha !} 
 \sim
\frac{(N_+/2)^{2u}}{(\prod_{\alpha=1}^{d_1}\lambda_\alpha^{p_\alpha}) N_+^u}
\frac{v^{2u}}{\prod_{\alpha=1}^{d_1} p_\alpha !} 
\sim O((N_+v^2)^u) \prod_{\alpha=1}^{d_1}(1/\lambda_\alpha)^{p_\alpha}. \label{A_term_inFC}
\ee
When some $\epsilon_\alpha$s are equal to $1$, the behavior of each term in (\ref{Jint3})  as $n_1^{(\alpha)}\sim N_+ \to \infty$ is obtained by multiplying equations (\ref{Cofbehave}) and (\ref{A_term_inFC}) as follows:
\be
 \left(\frac{d_1}{2d}\right)^{N_+} O(N_+^{N_-+a-(1+d_1)/2})O((N_+v^2)^u) \prod_{\alpha=1}^{d_1}(1/\lambda_\alpha)^{p_\alpha}. \label{evalJ}
\ee
Since each $(p_1,p_2,\dots,p_d)$ term of infinite series in (\ref{Jint3}) is obtained in the thermodynamic limit ($n^{(\alpha)}_1\sim L\to \infty$), $p_\alpha$s in the term should be regarded as finite values. So the absolute value of (\ref{evalJ}) decreases exponentially and becomes zero. Thus, only $J_{n_0,n_0,\ldots,n_0}$ survives on the right-hand side of (\ref{Jrs}), and the correlation functions for the DBC case $J_{\vec{r}\vec{s}}^{(\pm 1/2)}$ coincide with that for the PBC case, namely $G_{\vec{r}\vec{s}}^{(\pm 1/2)}$ in (\ref{GP3}).
\section{Conclusions and discussions}
For harmonic lattices, we have shown that the vacuum correlation function matrices $Q$ and $P$ for the dynamical variables and their conjugate momenta on sites are consistently expressed using Lauricella's C-type hypergeometric series $F_C^{(d)}$ in the thermodynamic limit, in any spatial dimension $d$ of the lattice. For lattices satisfying periodic boundary conditions, $Q_{\vec{r}\vec{s}}$ and $P_{\vec{r}\vec{s}}$ are given by (\ref{QP_L}) and (\ref{PPL}), and for those satisfying the Dirichlet boundary conditions, they are given by (\ref{QC_L}) and (\ref{PCL}). These hypergeometric series absolutely converge for spatial dimensions $d\geq 2$. Additionally, we have specifically observed that the correlation functions $Q_{\vec{r}\vec{s}}$ and $P_{\vec{r}\vec{s}}$ for sites near the center of the lattice under Dirichlet boundary conditions coincide with those in the lattice under periodic boundary conditions. 
One may examine analytically and numerically, as the sites move away from the boundary of lattice,  
how the correlation functions (26) and (27) converge to the corresponding expressions (18) and (19). 
We leave the detailed investigation of this behavior to future work.

In this paper, we have derived expressions for the correlation functions of $Q$ and $P$ in the thermodynamic limit for the relativistic case. 
The non-relativistic case for $Q$ has already been derived in \cite{IJ}. 
Their result (Theorem 2.1 in \cite{IJ}) is the same expression as our equation (\ref{QP_L}) except for the constant coefficients and the one-half discrepancy in the first two arguments of Lauricella's C-type hypergeometric function. The discrepancy arises from the difference between the non-relativistic Schr\"{o}dinger equation and the relativistic Klein-Gordon equation.

In the specific cases, it is known that Lauricella's C-type hypergeometric functions are factorized into the product of Gauss's hypergeometric function ${}_2F_1$s. 
As an example, let us consider $F_C^{(2)}$ appearing in the $Q$ of (\ref{QP_L}) for $d=2$ which is also referred as Appell's $F_4$ hypergeometric functions. 
For the massless harmonic lattice, the arguments of this hypergeometric functions are $v^2=1/4$ and it is factorized as follows\cite{Vidunas}}:
\begin{align}
& F_4\left(\frac{N+1/2}{2},\frac{N+3/2}{2};n_1+1,n_2+1;1/4,1/4\right) \nonumber \\
& ={}_{2}F_1\left(\frac{N+1/2}{2},\frac{N+3/2}{2};n_1+1;1/2\right){}_{2}F_1\left(\frac{N+1/2}{2},\frac{N+3/2}{2};n_2+1;1/2\right).
\label{F_Cfacts}
\end{align}
Further investigation into whether this expression is written in a more concise form, and whether any of the other correlation functions can be factorized in this manner, remains a future task.

Numerical calculations of $Q$ and $P$ matrices are essential for research using harmonic lattices on entanglement indices like R\'{e}nyi entropy and their corner functions, edge terms and contour functions, which represents the contribution to the indices from vertices, edges on the boundary of the physical system and from each site within it. 
However, as the dimensionality increases, creating programs to perform these computations quickly under memory constraints for sufficiently large-scale subsystems becomes increasingly difficult. The analytical solutions presented here, which express the correlation functions in $d$-dimensional space using hypergeometric functions, are expected to facilitate the creation of these programs. 

The following two issues remain: 1.We will investigate whether hypergeometric functions in correlation functions in this manuscript can be written in more concise forms as the factorization (\ref{F_Cfacts}). 2.For $d=2$ and $d=3$, we will obtain corner functions and edge terms for specific shapes and scales of subsystems based on the $Q$ and $P$ values computed by the program cord that employs expressions (\ref{QP_L}) and (\ref{PPL}). Then, we will compare these results with those from previous studies, which were computed based on the $Q$ and $P$ values in the finite-scale lattice of (\ref{corpq}) and (\ref{corpp}) or on those of integral expressions in (\ref{GP2}). 
\newpage\noindent


\begin{thebibliography}{99}
\bibitem{Amico}
L. Amico, R. Fazio, A. Osterloh, V.Vedral, Entanglement in many-body systems, 
Reviews of modern physics 80 (2), 517,(2008).
\bibitem{Nobili}
Cristiano De Nobili, Andre Coser and Erik Tonni,Journal of Statistical Mechanics: Theory and Experiment (2016) 083102, DOI 10.1088/1742-5468/2016/08/083102, 
arXiv:1604.02609.
\bibitem{Coser}
Andre Coser, Cristiano De Nobili and Erik Tonni, Journal of Physics A: Mathematical and Theoretical (2017) 50 314001, DOI 10.1088/1751-8121/aa7902,
arXiv:1701.08427. 
\bibitem{IK2MS_1}
M.Shimojo, S.Ishihara, H.Kataoka, A.Matsukawa, "Corner functions form entanglement indices of harmonic lattices", arXiv:2508.04992.
\bibitem{IJ} 
Ito, K., Jensen, A. "Hypergeometric Expression for the Resolvent of the Discrete Laplacian in Low Dimensions", Integr. Equ. Oper. Theory 93, 32 (2021), \\ https://doi.org/10.1007/s00020-021-02648-2, arXiv:2004.05866 [math-ph].
\bibitem{Vidunas}
Raimundas Vid$\bar{\text{u}}$nus, J.math. Anal. Appl.355(2009) 145-163.
\end{thebibliography}
\end{document}